%% file: plb.tex
\documentclass[aps,prl,twocolumn,showpacs,superscriptaddress,groupedaddress]{revtex4}  

\usepackage{graphicx}  

\usepackage{dcolumn}   
\usepackage{bm}        
\usepackage{amssymb}   
\usepackage{amsmath}   

\RequirePackage{color}\definecolor{RED}{rgb}{1,0,0}\definecolor{BLUE}{rgb}{0,0,1} \definecolor{PURPLE}{rgb}{1,0,1}

\newcommand{\fb}{\ensuremath{\mathrm{fb}^{-1}}}
\newcommand{\Eslash}{\mbox{$ E \kern-0.6em\slash$                }}
\newcommand{\pslash}{\mbox{$ p \kern-0.45em\slash$                }}
\newcommand{\etmiss}{\mbox{$\Eslash_T\!$                        }}
\newcommand{\ptmiss}{\mbox{$\pslash_T\!$                        }}

\newcommand{\ttbar}{\mbox{$t{\bar t}~$}}

\def\MET{{\mbox{$E\kern-0.57em\raise0.19ex\hbox{/}_{T}$}}}
\def\met{{\mbox{$E\kern-0.57em\raise0.19ex\hbox{/}_{T}$}}}

\def\ifb{~fb$^{-1}$}

\def\ppbar{$p\bar{p}$}

\def\lmet{$WH\rightarrow \ell\kern-0.45em\raise0.19ex\hbox{/} \nu b\bar{b}$}

\begin{document}

\hspace{5.2in} \mbox{FERMILAB-PUB-12-541-E}

\title{Measurement of the $\bm{ p\bar{p} \to W+b+X}$ production cross section at $\bm{ \sqrt{s}=1.96}$~TeV}
\input author_list.tex 
\begin{abstract}

We present a measurement of the cross section for $W$ boson production in association with at least one \mbox{$b$-quark} jet in 
proton-antiproton collisions.
The measurement is made using  data corresponding to an integrated luminosity of $6.1$\ifb~recorded with the D0 detector 
at the Fermilab Tevatron \ppbar~Collider at $\sqrt{s}=1.96$~TeV. We measure an inclusive cross section of \mbox{$\sigma(W~(\to\mu\nu) + b + X) = 1.04~\pm~0.05\thinspace$(stat.)~$\pm~0.12 \thinspace$(syst.)~pb} and $\sigma(W (\to e\nu) + b + X) = 1.00$~$\pm~0.04 \thinspace$(stat.)~$\pm~0.12 \thinspace$(syst.)~pb in the phase space defined by $p_T^\nu > 25$~GeV, $p_T^{\text{$b$-jet}}>20$~GeV, $|\eta^{\text{$b$-jet}}|<1.1$, and a muon (electron) with $p_T^\ell>20$~GeV and  $|\eta^\mu|<1.7$ ($|\eta^e|<1.1$ or $1.5<|\eta^e|<2.5$). 
The combined result per lepton family is $\sigma(W (\to \ell \nu) + b + X) = 1.05$~$\pm~0.12 \thinspace$(stat.+syst.) for $|\eta^\ell|<1.7$. The results are in agreement with predictions from next-to-leading order QCD calculations using \textsc{mcfm}, $\sigma(W+b)\cdot  {\cal B}(W \to \ell \nu)= 1.34~^{+0.41}_{-0.34}\thinspace(\textrm{syst.})$, and also with predictions from the \textsc{sherpa} and \textsc{madgraph} Monte Carlo event generators.
\end
{abstract}

\pacs{12.38.Qk, 13.85.Qk, 14.65.Fy, 14.70.Fm}
\maketitle

The measurement of the production cross section of a $W$ boson in association with a $b$-quark jet provides a stringent test of quantum chromodynamics (QCD). Processes involving $W/Z$ bosons in association with $b$ quarks are also the largest backgrounds in studies of the standard model (SM) Higgs boson decaying to two $b$ quarks, in measurements of top quark properties in both single and pair production, and in numerous searches for physics beyond the SM. The cross section for the process \mbox{$p\bar{p}\to W+b+X$} has been calculated with next-to-leading order (NLO) precision~\cite{assoc_wb1, assoc_wb}.
Subprocesses at NLO  include $q\bar{q}~\to~Wb\bar{b},~q\bar{q}~\to~Wb\bar{b}g$, and $qg~\to~Wb\bar{b}q'$. An additional small contribution comes from sea $b$ quarks in the incoming proton or antiproton, $bq \to W bq'$.

In this letter we describe a measurement of the cross section for $W$ boson production in association with $b$-quark jets in \ppbar~interactions, where a $W$ boson is identified via its electronic or muonic decay modes. A measurement of $W+b$ production cross section with up to two jets at $\sqrt{s}=1.96$~TeV has been published by the CDF Collaboration~\cite{wbb_cdf} and an inclusive measurement has been published by the ATLAS Collaboration~\cite{wbb_atlas} at $\sqrt{s}=7$~TeV. 
The measured production cross section reported by CDF is \mbox{$ \sigma \cdot  {\cal B}(W \to \ell \nu)  = 2.74 \pm 0.27\thinspace(\text{stat.})\pm 0.42 \thinspace(\text{syst.})$}~pb ($\ell=e,~\nu$), while the theoretical expectation for this quantity based on NLO calculations is \mbox{$1.22 \pm 0.14 \thinspace \text{(syst.)}$~pb}~\cite{wbb_cdf}. With the CDF measurement of $W+b$ production exceeding significantly the NLO prediction, while the ATLAS result is in agreement with the expectation, an independent measurement is important to understand the production of $W$ bosons in association with $b$ jets at hadron colliders.

The data used in this analysis were collected between July 2006 and December 2010 using the D0 detector at the Fermilab
Tevatron Collider at $\sqrt{s}~=~1.96$~TeV, and correspond to an integrated luminosity of 6.1~\fb. 
We first briefly describe the main components of the D0 Run II detector
\cite{run2det} relevant to this analysis. The D0 detector has a central tracking system consisting of a 
silicon microstrip tracker (SMT) \cite{layer0} and a central fiber tracker (CFT), 
both located within a 2~T superconducting solenoidal 
magnet, with designs optimized for tracking and 
vertexing at pseudorapidities $|\eta|<3$ and $|\eta|<2.5$, respectively~\cite{coord}.
A liquid argon and uranium calorimeter has a 
central section (CC) covering pseudorapidities $|\eta| \lesssim 1.1$, and two end calorimeters (EC) that extend coverage 
to $|\eta|\approx 4.2$, with all three housed in separate 
cryostats~\cite{calopaper}. An outer muon system, at $|\eta|<2$, 
consists of a layer of tracking detectors and scintillation trigger 
counters in front of 1.8~T toroids, followed by two similar layers 
after the toroids. 
Luminosity is measured using plastic scintillator 
arrays located in front of the EC cryostats. The trigger and data 
acquisition systems are designed to accommodate the high instantaneous luminosities 
of Run II.

The $W+b$ candidates are selected by triggering on single lepton or lepton-plus-jet signatures with a three-level trigger system. 
The trigger efficiencies are approximately $70\%$ for the muon channel and $95\%$ for the electron channel. 

$W$ boson candidates are identified in the $\mu+\nu$ and $e+\nu$ decay channels whereas a small fraction of selected events arises from leptonical decaying tau leptons.
Offline event selection requires a reconstructed primary $p\bar{p}$~interaction primary vertex (PV) that has at least three associated tracks and is located within $60$~cm of the center of the detector along the beam direction.  The vertex selection for $W+b$ events is about $97\%$ efficient as measured in simulations.

Electrons are identified using calorimeter and tracking information. The selection requires exactly one electron with transverse momentum $p_T^e>20$~GeV identified by an electromagnetic (EM) shower in the central ($|\eta^e|<1.1$) or endcap ($1.5 < |\eta^e| < 2.5$) calorimeter by comparing the longitudinal and transverse shower profiles to those of simulated electrons. 
The showers must be spatially isolated from other energetic particles, deposit most of their energy in the EM part of the calorimeter, and pass a likelihood criterion that includes a spatial track match. In the central detector region, an $E/p$ requirement is applied, where $E$ is the energy of the calorimeter cluster and $p$ is the momentum of the track. The transverse momentum measurement of electrons is based on calorimeter energy information.

The muon selection requires the candidate to be reconstructed from hits in the muon system and matched to a reconstructed track in the
central tracker. The transverse momentum of the muon must exceed $p_T^{\mu}>20$~GeV, with  $|\eta^{\mu}|<1.7$. 
Muons are required to be spatially isolated from other energetic particles using information from the central tracking detectors and calorimeter~\cite{diboson_prd}. Muons from cosmic rays are rejected by applying a timing criterion on the hits in the scintillator layers and by applying restrictions on the displacement of the muon track with respect to the selected PV.

Candidate $W+\textrm{jets}$ events are then selected by requiring at least one reconstructed jet with $|\eta^{\text{jet}}| < 1.1$ and $p_T^{\text{jet}} > 20$~GeV. Jets are reconstructed from energy deposits in the calorimeter using the iterative midpoint cone algorithm~\cite{jet_algo} and a cone of radius $\Delta R=0.5$ in $y$-$\varphi$ space~\cite{coord}. The energies of jets are corrected for detector response, the presence of noise and multiple \ppbar~interactions, and for energy deposited outside of the jet reconstruction cone. To enrich the sample with $W$ bosons, events are required to have missing transverse energy \etmiss $>25$~GeV due to the neutrino escaping detection.

Background processes for this analysis are electroweak $W+\textrm{jets}/\gamma$ production, $Z/\gamma^*$ production, $t{\bar t}$ and single top quark production, diboson production, and multijet events with jets misidentified as leptons. The $W+b$ signal and SM background processes are simulated using a combination of \textsc{pythia \textrm{v}6.409}~\cite{pythia} and \textsc{alpgen \textrm{v}2.3}~\cite{alpgen} with \textsc{pythia} providing parton showering and hadronization. We use \textsc{pythia} Tune A with CTEQ6L1~\cite{pdf_cteq6M} parton distribution functions (PDFs) and perform a detailed \textsc{geant}-based~\cite{geant} simulation of the D0 detector. 
The $V{\rm +jets}$  ($V=W/Z$) processes are normalized to the inclusive $W$ and $Z$-boson cross sections calculated at NNLO~\cite{hamberg}.
The $Z$-boson $p_T$ distribution is modeled to match the distribution observed in data~\cite{zpt_xsec}, taking into account the dependence on the number of reconstructed jets. To reproduce the $W$-boson $p_T$ distribution in simulated events, the product of the measured $Z$-boson $p_T$ spectrum and the ratio of $W$ to $Z$-boson $p_T$\ distributions at NLO is used as correction.  NLO+NNLL (next-to-next-to-leading log) calculations are used to normalize \ttbar\ production~\cite{mochuwer}, while single top quark production is normalized to NNLO~\cite{single_top}. The NLO $WW$, $WZ$, and $ZZ$ production cross section values are obtained with \textsc{mcfm} program~\cite{mcfm}. For the $W$+heavy-flavor jet ($b$ or $c$ quark) events, the ratio of the \textsc{alpgen} prediction to the \textsc{NLO} prediction for $W+b\bar{b}$ and ~$W+c\bar{c}$ is obtained from \textsc{mcfm}~\cite{mcfm} and applied as a correction factor.  The simulation is also corrected for the trigger efficiencies measured in data.

Instrumental backgrounds and those from semileptonic decays of hadrons, referred to  as ``multijet'' background, are estimated from data. The
instrumental background is important for the electron channel, where a jet with a high electromagnetic fraction can pass electron identification criteria, or a photon can be
misidentified as an electron. In the muon channel, the multijet background is less significant and arises mainly from the semileptonic decay of heavy quarks in which the muon satisfies the isolation requirements. We require that the $W$ boson candidates have a transverse mass $M_T$~\cite{mtw} satisfying \mbox{$40~\text{GeV}+~\frac{1}{2}\etmiss < M_T < 120$~GeV} to suppress multijet background and mis-reconstructed events. The average efficiency determined in simulation for a $W+b$ signal to pass these requirements is about $82\%$.

Identification of $b$ jets is crucial for this measurement.
Once the inclusive $W+{\rm jets}$ sample is defined, the jets considered for $b$ tagging are subject to a requirement called taggability. This requirement is imposed to decouple the performance of the $b$-jet identification from detector effects. For a jet to be taggable, it must contain at least two tracks with at least one hit in the SMT, $p_T>1$~GeV for the highest-$p_T$ track and $p_T>0.5$~GeV for the next-to-highest $p_T$ track. The efficiency for a jet to be taggable is about $90\%$ in the selected phase space.

\begin{figure}[h!]
    \begin{minipage}{0.49\textwidth}
      \centering 
      \includegraphics[scale=0.467]{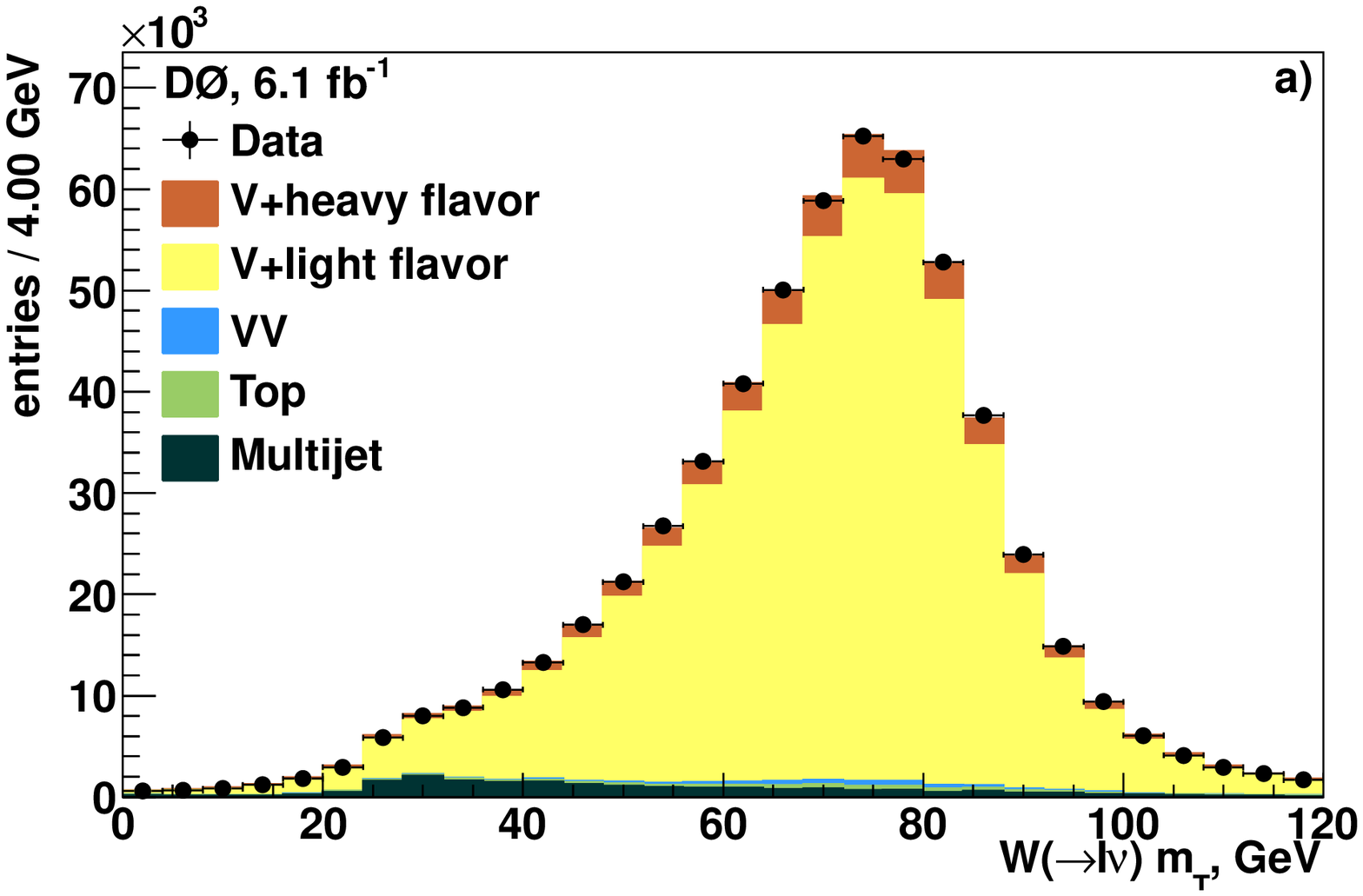}
      \end{minipage}
    \hfill
    \begin{minipage}{0.49\textwidth}
      \centering 
      \includegraphics[scale=0.467]{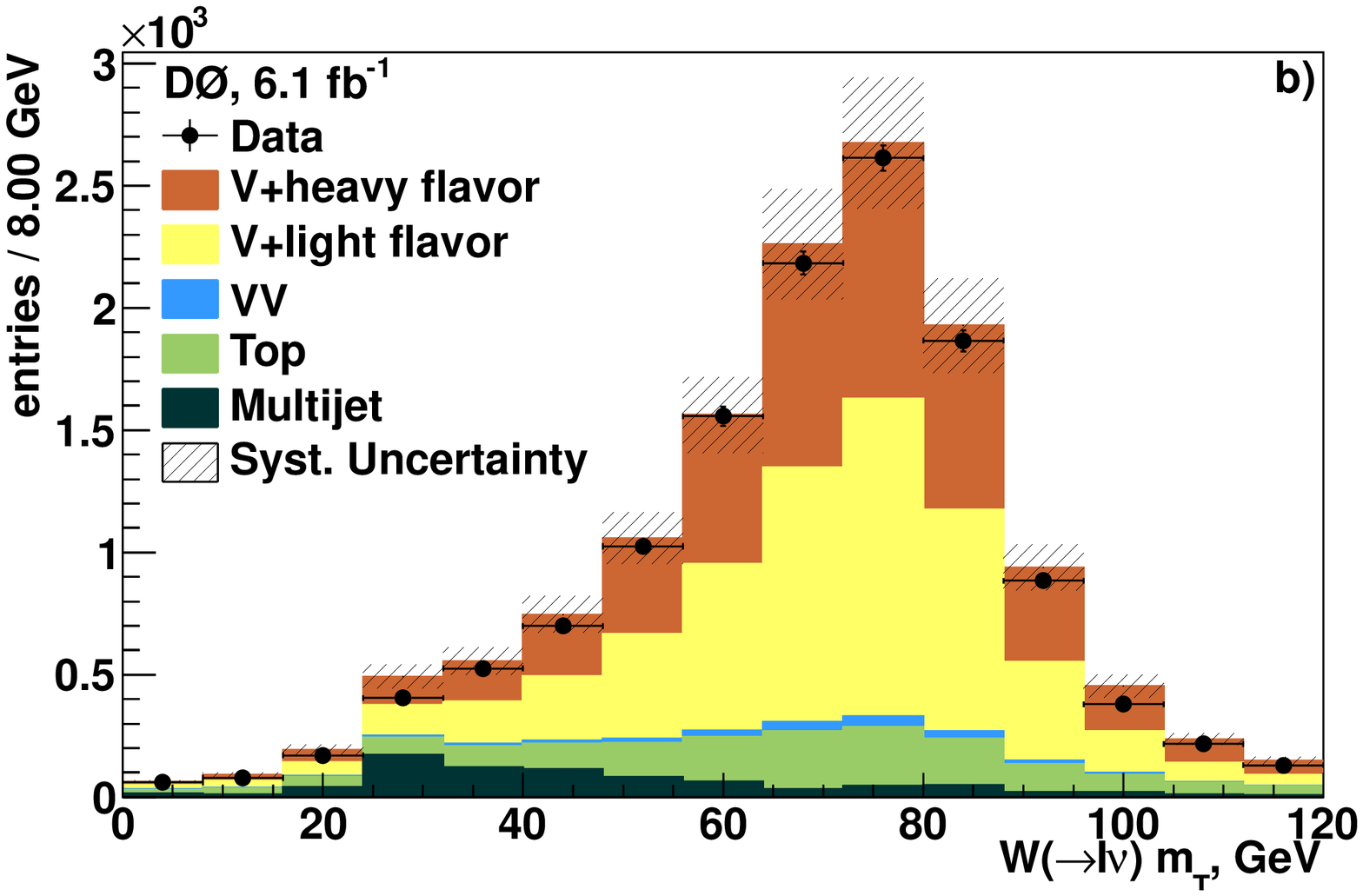}
    \end{minipage}
    \caption{\footnotesize [color online] Transverse mass of the $\ell\nu$ system (a) before and (b) after  $b$-jet identification. The data are shown by black markers, simulated background processes are shown by filled histograms. The data uncertainties are statistical only. An estimate of the systematic uncertainty on the simulated background processes is shown by the shaded bands
     \label{fig:wm}}
\end{figure}

 The D0 $b$-tagging algorithm for identifying heavy flavor jets is based on a combination of variables sensitive to the presence secondary vertices (SV) or tracks displaced from the PV. This analysis uses an updated $b$ tagger utilizing a multivariate analysis (MVA)~\cite{tmva, gamma_b} that provides improved performance over the previous neural network based algorithm~\cite{bid_nim}.   
The most sensitive input variables to the MVA are the number of reconstructed secondary vertices in the jet, the invariant mass of charged particles associated with the SV ($M_{\text{SV}}$), the number of tracks used to reconstruct the SV, the two-dimensional decay length significance of the SV in the plane transverse to the beam, a weighted combination of the tracks' transverse impact parameter significances, and the probability that the tracks from the jet originate from the PV, which is referred to as the jet lifetime probability (JLIP). The MVA provides a continuous output value that tends towards one for $b$ jets and zero for non-$b$ jets.
Events are considered in which at least one jet passes a tight MVA requirement corresponding to an efficiency of $\approx 50\%$ for $b$ jets. The likelihood for a light jet ($u$, $d$, $s$ quarks and gluons) to be misidentified for the corresponding MVA selection is about $0.5$\%. 
Simulated events are corrected to have the same efficiencies for taggability and $b$-tagging requirements as found in data. These corrections are derived in a flavor dependent manner ~\cite{bid_nim}, using independent $QCD$ enriched data samples and simulated events with enriched light and heavy jet contributions. Jets containing $b$ quarks have a different energy response and receive an additional energy correction of about $6\%$ as determined from simulation. Figure~\ref{fig:wm} shows the transverse mass of the candidate events before and after applying $b$-jet identification.

In addition to the MVA output, we perform further selections using $M_{\text{SV}}$ and JLIP variables.
$M_{\text{SV}}$ provides good discrimination between $b$, $c$, and light quark jets due to their different masses~\cite{gamma_b}.
The two variables together take into account the kinematics of the event and, in order to further improve the separation power, they are
combined in a single variable ${\cal D}_{\text{MJL}}~=\frac{1}{2}~\left(M_{\text{SV}}/(5~\text{GeV}) - \ln(\text{JLIP})/20 \right )$~\cite{zbzlf}.
A loose criterion for an event to pass at least ${\cal D}_{\text{MJL}}>0.1$ is applied to remove poorly reconstructed events. The efficiency for signal events to pass this selection is about $97\%$.

The numbers of expected and observed events before and after applying the $b$-jet identification in data and simulation are listed in Table~\ref{tab:yields}. The $b$-tagging column includes the selection requirement on ${\cal D}_{\text{MJL}}$.

\begin{table}[ht]
  \begin{tabular}{l|rrl|rll}
    \hline \hline
    Process            &  \multicolumn{3}{|c|}{No $b$-tag} & \multicolumn{3}{c}{$b$-tag}                 \\ \hline
    $V+$heavy flavor   & 41093    & $\pm$ & 8924   & 5068  & $\pm$ & 1124   \\ \hline
    $V+$light flavor   & 516661   & $\pm$ & 56734  & 5718  & $\pm$ & 678    \\ \hline
    Diboson            & 4728     & $\pm$ & 519    & 222   & $\pm$ & 26     \\ \hline
    Top                & 5431     & $\pm$ & 536    & 1602  & $\pm$ & 181    \\ \hline
    Multijet           & 20527    & $\pm$ & 4458   & 794   & $\pm$ & 180    \\ \hline
    Expected events    & 588440   & $\pm$ & 57610  & 13405 & $\pm$ & 1338  \\ \hline
    Data            &  \multicolumn{3}{|c|}{586289}& \multicolumn{3}{c}{12793}                 \\ \hline
    \hline 
  \end{tabular}
  \caption{Numbers of events for data and contributing processes before and after applying $b$-jet identification. Uncertainties include statistical and systematic
contributions. The contribution of $Z+\textrm{jets}$ events to the $V+\textrm{jets}$ samples is $\approx 5\%$ for heavy and light flavor jets before and after $b$-tagging. \label{tab:yields}}
\end{table}

We measure the fraction of $W+b+X$ events in the final selected sample by performing a binned maximum likelihood fit 
to the observed data distribution of the ${\cal D}_{\text{MJL}}$ discriminant in our sample shown in Fig.~\ref{fig:templates}. The templates for $W$+light flavor, $W+b$, and $W+c$ jets shown in Fig.~\ref{fig:templates}
are taken from the efficiency-corrected simulation. Expected contributions from $Z$+jets, single top quark, $t\bar{t}$, diboson, and multijet production 
are subtracted from the data. After performing the fits, we obtain the number of events with different jet flavors listed in Table~\ref{tab:fit_wbb_frac}.

The measured cross sections are presented at the particle level by correcting for detector acceptance, selection-efficiencies, and $b$-jet identification. We quote our result as a cross section in a restricted phase space: at least one $b$-jet with  $p_T^{\text{$b$-jet}}>20$~GeV, $|\eta^{\text{$b$-jet}}|<1.1$ and a muon with $p_T^\mu~>~20$~GeV and $|\eta^\mu|<1.7$ or an electron with $p_T^e>20$~GeV and $|\eta^e|<1.1$ or $1.5<|\eta^e|<2.5$. For the neutrino momentum we require $p_T^\nu>25$~GeV.

\begin{table}[ht]
  \begin{tabular}{l|lc|lc}
    \hline \hline
                                &  \multicolumn{2}{c|}{$W\to \mu \nu$}  & \multicolumn{2}{c}{$W\to e \nu$} \\ \hline
    Process\phantom{xxxx}        &   \multicolumn{1}{l}{Events}        & Fraction   & \multicolumn{1}{l}{Events}         & Fraction       \\ \hline
    $W+b$                        &   $1306 \pm 166$     & $0.3\pm0.04$    & $1676 \pm 212$ & $0.27\pm0.03$  \\ \hline
    $W+c$                        &   $664  \pm 97 $     & $0.1\pm0.02$    & $1096 \pm 159$ & $0.18\pm0.03$\\ \hline
    $W+\textrm{l.f.}$            &   $2152 \pm 265$     & $0.5\pm0.07$    & $3479 \pm 425$ & $0.56\pm0.07$\\ \hline
    Data$-$Bkgd                  &   $4127 \pm 150$     &                 & $6255 \pm 168$ &      \\ \hline
    \hline 
  \end{tabular}
  \caption{Estimated numbers of $W+\text{jet}$ events from fitting the flavor-specific processes, along with the expected background of $W$ boson processes and the data after subtracting $Z$+jets, single top quark, $t\bar{t}$, and diboson background processes. $l.f.$ stands for light flavor jets. Uncertainties include statistical and systematic contributions. \label{tab:fit_wbb_frac}}
\end{table}

Systematic uncertainties are determined by varying experimental parameters and efficiency/acceptance corrections by one standard deviation and propagating the effect on ${\cal D}_{\text{MJL}}$. The systematic uncertainties are dominated by effects related to the measurement of jets.  The contributions from jet energy resolution, jet modeling, and detector effects are about $2.5\%,~3\%$,~and~$4\%$, respectively. Uncertainties on $b$-jet identification are determined in data and simulations by using $b$-jet-enriched samples and are about $2\%-5\%$ per jet. The uncertainties due to lepton identification are about $2\%$. The integrated luminosity is known to a precision of $6.1\%$~\cite{lumi}. The uncertainty of the template fit is estimated by varying the normalization and shape from the data corrections of the $W$ boson processes and the fit parameters (about $6\%$). By summing the uncertainties in quadrature we obtain a final total systematic uncertainty on the cross section measurements of approximately $12\%$.

The cross section times branching fraction is calculated by dividing the number of signal events measured by integrated luminosity ($\mathcal{L}$), acceptance ($\mathcal{A}$), and efficiencies ($\epsilon$) of the selection requirements:

\begin{figure}[h!]

    \begin{minipage}{0.49\textwidth}
      \centering 
      \includegraphics[scale=0.45]{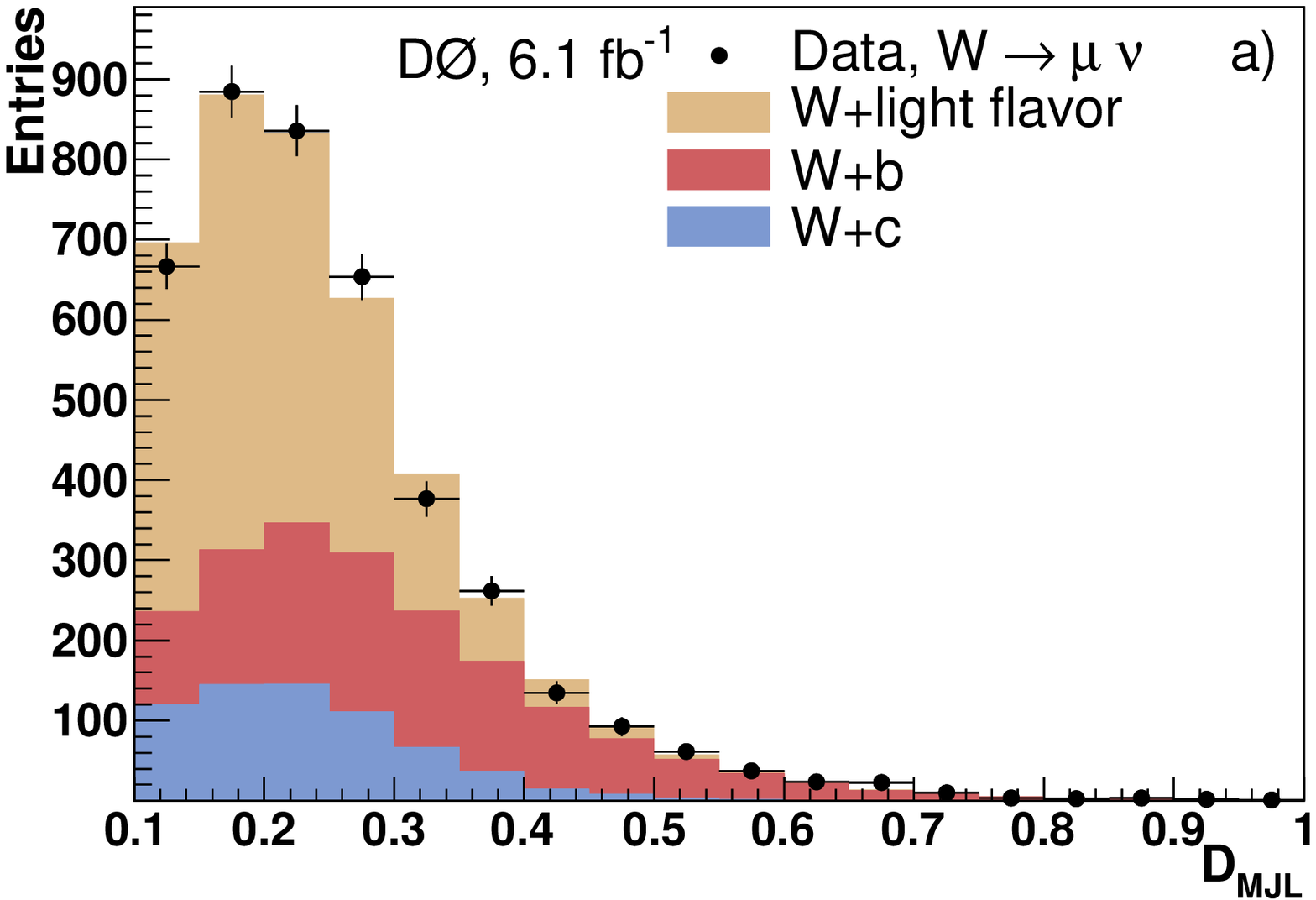}
      \end{minipage}
    \hfill
    \begin{minipage}{0.49\textwidth}
      \centering 
      \includegraphics[scale=0.45]{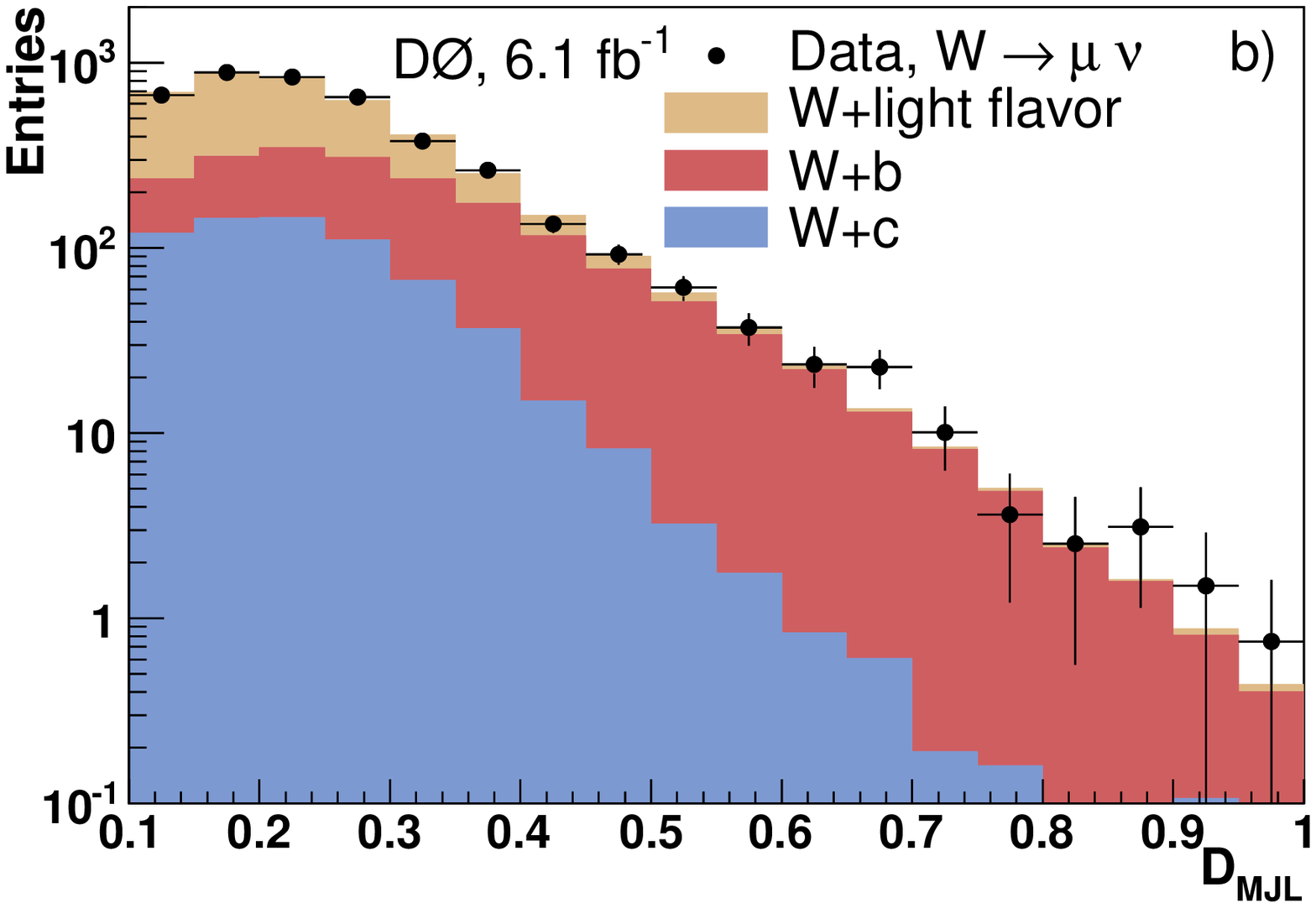}
    \end{minipage}
    \caption{\footnotesize
      [color online] Contributions of the various jet flavors normalized to the measured cross section obtained from a fit in the $W \to \mu \nu$ channel on both (a) linear and (b) logarithmic scales.  The various $W+\textrm{jets}$ processes are shown as filled histograms and data, after the subtraction of contributions from Drell-Yan, diboson, and top quark production, are represented with black markers. The uncertainties include both statistical and systematic contributions. 
    \label{fig:templates}}
\end{figure}

\begin{equation}
  \sigma(W+b) \cdot  {\cal B}(W \to \ell \nu)  = \frac{N_{W+b} }{{\cal L} \cdot {\cal A} \cdot  \epsilon},
\label{eq:qcd_xsec}
\end{equation}
where $\epsilon$ is given by the product of the trigger, object reconstruction, and selection efficiencies.

We first present results separately for the muon channel and electron channel because they are performed in slightly different requirements on the phase space of the lepton and then combine using a common phase space. We measure from the cross section in the muon channel where $W \to \mu \nu$ in a visible phase space defined by $p_T^{\mu}>20$~GeV, $|\eta^\mu|<1.7$ with at least one $b$-jet limited to $p_T^{\text{$b$-jet}}>20$~GeV and $|\eta^{\text{$b$-jet}}| < 1.1$ as,

\begin{equation}
  \begin{split}
    \sigma(W+b) \cdot  {\cal B}(W \to \mu \nu)  = \phantom{xxxxxx}\\
    1.04\pm 0.05 \thinspace \textrm{(stat.)} \pm 0.12 \thinspace \textrm{(syst.)~pb.}  
  \end{split}
\end{equation}

\noindent
We perform an NLO QCD prediction using \textsc{mcfm} v6.1, based on CTEQ6M PDF~\cite{pdf_cteq6M} and a central scale of $M_W +2m_b$, where $m_b = 4.7$~GeV is the mass of the $b$ quark. Uncertainties are estimated by varying renormalization and factorization scales by a factor of two in each direction, varying $m_b$ between $4.2$ and $5$~GeV, and by using an alternative PDF set.  The \textsc{mcfm} calculation predicts $\sigma(W+b)\cdot  {\cal B}(W \to \mu \nu)= 1.34~^{+0.40}_{-0.33}~(\textrm{scale}) \pm 0.06~(\textrm{PDF})~^{+0.09}_{-0.05}~(m_b)$~pb.
Predictions obtained using \textsc{sherpa} v1.4 and CTEQ6.6 PDFs~\cite{pdf_cteq6M} lead to a value $1.21 \pm 0.03 \thinspace(\textrm{stat.})$~pb. Using \textsc{madgraph5}~\cite{mg5} with CTEQ6L1 PDFs, we obtain $1.52 \pm 0.02 \thinspace(\textrm{stat.})$~pb. Uncertainties for scale variations, PDFs, and the $b$-quark mass are on the order of about 30\%.

In the electron channel, we measure the cross section times branching fraction by selecting $p_T^{e}>20$~GeV, $|\eta^e|<1.1$ or $1.5<|\eta^e|<2.5$, at least one $b$-jet as above and obtain

\begin{equation}
  \begin{split}
    \sigma(W+b) \cdot  {\cal B}(W \to e \nu)  = \phantom{xxxxxx}\\
    1.00\pm 0.04 \thinspace \textrm{(stat.)} \pm 0.12 \thinspace \textrm{(syst.)~pb.}
  \end{split}
\end{equation}

\noindent
The \textsc{mcfm} calculated cross section for this channel is $\sigma(W+b)\cdot  {\cal B}(W\to e \nu)=1.28~^{+0.40}_{-0.33}~(\textrm{scale}) \pm 0.06~(\textrm{PDF})~^{+0.09}_{-0.05}~(m_b)$~pb. The \textsc{sherpa} prediction is $1.08 \pm 0.03 \thinspace (\textrm{stat.})$~pb, while the \textsc{madgraph5} prediction is $1.44\pm0.02 \thinspace (\textrm{stat.})$~pb. The combined systematic effect scale, PDF and $m_b$ variations is also around 30\%.

Using the \textsc{mcfm} prediction we extrapolate the measurement in the electron final state to the same selection requirements as the muon final state to allow for a consistent combination. Combining the results in $W \to \mu \nu$ and $W \to e\nu$ decays we obtain 

\begin{equation}
  \begin{split}
    \sigma(W+b) \cdot  {\cal B}(W \to \ell \nu)  = \phantom{xxxxxx}\\
    1.05\pm 0.03 \thinspace \textrm{(stat.)} \pm 0.12 \thinspace \textrm{(syst.)~pb.}
  \end{split}
\end{equation}

The small experimental uncertainty should allow to further constrain theoretical predictions. 
In summary, we have performed a measurement of the inclusive cross section for $W$ boson production in association with at least one $b$-jet at $\sqrt{s}=1.96$~TeV, considering final states with $W \to \mu\nu$ ($W \to e \nu$) events in a restricted phase space of $p_T^{\ell}>20$~GeV, $|\eta^\mu|<1.7$ ($|\eta^e|<1.1$ or $1.5<|\eta^e|<2.5$), with $b$ jets limited to $p_T^{\text{$b$-jet}}>20$~GeV and $|\eta^{\text{$b$-jet}}| < 1.1$.  
The measured cross sections agree within uncertainties with NLO QCD calculations and predictions obtained using the \textsc{sherpa} and \textsc{madgraph} generators.

\input acknowledgement.tex   

\end{document}
%

%% file: author_list.tex
\affiliation{LAFEX, Centro Brasileiro de Pesquisas F\'{i}sicas, Rio de Janeiro, Brazil}
\affiliation{Universidade do Estado do Rio de Janeiro, Rio de Janeiro, Brazil}
\affiliation{Universidade Federal do ABC, Santo Andr\'e, Brazil}
\affiliation{University of Science and Technology of China, Hefei, People's Republic of China}
\affiliation{Universidad de los Andes, Bogot\'a, Colombia}
\affiliation{Charles University, Faculty of Mathematics and Physics, Center for Particle Physics, Prague, Czech Republic}
\affiliation{Czech Technical University in Prague, Prague, Czech Republic}
\affiliation{Center for Particle Physics, Institute of Physics, Academy of Sciences of the Czech Republic, Prague, Czech Republic}
\affiliation{Universidad San Francisco de Quito, Quito, Ecuador}
\affiliation{LPC, Universit\'e Blaise Pascal, CNRS/IN2P3, Clermont, France}
\affiliation{LPSC, Universit\'e Joseph Fourier Grenoble 1, CNRS/IN2P3, Institut National Polytechnique de Grenoble, Grenoble, France}
\affiliation{CPPM, Aix-Marseille Universit\'e, CNRS/IN2P3, Marseille, France}
\affiliation{LAL, Universit\'e Paris-Sud, CNRS/IN2P3, Orsay, France}
\affiliation{LPNHE, Universit\'es Paris VI and VII, CNRS/IN2P3, Paris, France}
\affiliation{CEA, Irfu, SPP, Saclay, France}
\affiliation{IPHC, Universit\'e de Strasbourg, CNRS/IN2P3, Strasbourg, France}
\affiliation{IPNL, Universit\'e Lyon 1, CNRS/IN2P3, Villeurbanne, France and Universit\'e de Lyon, Lyon, France}
\affiliation{III. Physikalisches Institut A, RWTH Aachen University, Aachen, Germany}
\affiliation{Physikalisches Institut, Universit\"at Freiburg, Freiburg, Germany}
\affiliation{II. Physikalisches Institut, Georg-August-Universit\"at G\"ottingen, G\"ottingen, Germany}
\affiliation{Institut f\"ur Physik, Universit\"at Mainz, Mainz, Germany}
\affiliation{Ludwig-Maximilians-Universit\"at M\"unchen, M\"unchen, Germany}
\affiliation{Fachbereich Physik, Bergische Universit\"at Wuppertal, Wuppertal, Germany}
\affiliation{Panjab University, Chandigarh, India}
\affiliation{Delhi University, Delhi, India}
\affiliation{Tata Institute of Fundamental Research, Mumbai, India}
\affiliation{University College Dublin, Dublin, Ireland}
\affiliation{Korea Detector Laboratory, Korea University, Seoul, Korea}
\affiliation{CINVESTAV, Mexico City, Mexico}
\affiliation{Nikhef, Science Park, Amsterdam, the Netherlands}
\affiliation{Radboud University Nijmegen, Nijmegen, the Netherlands}
\affiliation{Joint Institute for Nuclear Research, Dubna, Russia}
\affiliation{Institute for Theoretical and Experimental Physics, Moscow, Russia}
\affiliation{Moscow State University, Moscow, Russia}
\affiliation{Institute for High Energy Physics, Protvino, Russia}
\affiliation{Petersburg Nuclear Physics Institute, St. Petersburg, Russia}
\affiliation{Instituci\'{o} Catalana de Recerca i Estudis Avan\c{c}ats (ICREA) and Institut de F\'{i}sica d'Altes Energies (IFAE), Barcelona, Spain}
\affiliation{Uppsala University, Uppsala, Sweden}
\affiliation{Lancaster University, Lancaster LA1 4YB, United Kingdom}
\affiliation{Imperial College London, London SW7 2AZ, United Kingdom}
\affiliation{The University of Manchester, Manchester M13 9PL, United Kingdom}
\affiliation{University of Arizona, Tucson, Arizona 85721, USA}
\affiliation{University of California Riverside, Riverside, California 92521, USA}
\affiliation{Florida State University, Tallahassee, Florida 32306, USA}
\affiliation{Fermi National Accelerator Laboratory, Batavia, Illinois 60510, USA}
\affiliation{University of Illinois at Chicago, Chicago, Illinois 60607, USA}
\affiliation{Northern Illinois University, DeKalb, Illinois 60115, USA}
\affiliation{Northwestern University, Evanston, Illinois 60208, USA}
\affiliation{Indiana University, Bloomington, Indiana 47405, USA}
\affiliation{Purdue University Calumet, Hammond, Indiana 46323, USA}
\affiliation{University of Notre Dame, Notre Dame, Indiana 46556, USA}
\affiliation{Iowa State University, Ames, Iowa 50011, USA}
\affiliation{University of Kansas, Lawrence, Kansas 66045, USA}
\affiliation{Kansas State University, Manhattan, Kansas 66506, USA}
\affiliation{Louisiana Tech University, Ruston, Louisiana 71272, USA}
\affiliation{Northeastern University, Boston, Massachusetts 02115, USA}
\affiliation{University of Michigan, Ann Arbor, Michigan 48109, USA}
\affiliation{Michigan State University, East Lansing, Michigan 48824, USA}
\affiliation{University of Mississippi, University, Mississippi 38677, USA}
\affiliation{University of Nebraska, Lincoln, Nebraska 68588, USA}
\affiliation{Rutgers University, Piscataway, New Jersey 08855, USA}
\affiliation{Princeton University, Princeton, New Jersey 08544, USA}
\affiliation{State University of New York, Buffalo, New York 14260, USA}
\affiliation{University of Rochester, Rochester, New York 14627, USA}
\affiliation{State University of New York, Stony Brook, New York 11794, USA}
\affiliation{Brookhaven National Laboratory, Upton, New York 11973, USA}
\affiliation{Langston University, Langston, Oklahoma 73050, USA}
\affiliation{University of Oklahoma, Norman, Oklahoma 73019, USA}
\affiliation{Oklahoma State University, Stillwater, Oklahoma 74078, USA}
\affiliation{Brown University, Providence, Rhode Island 02912, USA}
\affiliation{University of Texas, Arlington, Texas 76019, USA}
\affiliation{Southern Methodist University, Dallas, Texas 75275, USA}
\affiliation{Rice University, Houston, Texas 77005, USA}
\affiliation{University of Virginia, Charlottesville, Virginia 22904, USA}
\affiliation{University of Washington, Seattle, Washington 98195, USA}
\author{V.M.~Abazov} \affiliation{Joint Institute for Nuclear Research, Dubna, Russia}
\author{B.~Abbott} \affiliation{University of Oklahoma, Norman, Oklahoma 73019, USA}
\author{B.S.~Acharya} \affiliation{Tata Institute of Fundamental Research, Mumbai, India}
\author{M.~Adams} \affiliation{University of Illinois at Chicago, Chicago, Illinois 60607, USA}
\author{T.~Adams} \affiliation{Florida State University, Tallahassee, Florida 32306, USA}
\author{G.D.~Alexeev} \affiliation{Joint Institute for Nuclear Research, Dubna, Russia}
\author{G.~Alkhazov} \affiliation{Petersburg Nuclear Physics Institute, St. Petersburg, Russia}
\author{A.~Alton$^{a}$} \affiliation{University of Michigan, Ann Arbor, Michigan 48109, USA}
\author{A.~Askew} \affiliation{Florida State University, Tallahassee, Florida 32306, USA}
\author{S.~Atkins} \affiliation{Louisiana Tech University, Ruston, Louisiana 71272, USA}
% ...
%
%
\collaboration{The D0 Collaboration\footnote{with visitors from
$^{a}$Augustana College, Sioux Falls, SD, USA,
$^{b}$The University of Liverpool, Liverpool, UK,
$^{c}$UPIITA-IPN, Mexico City, Mexico,
$^{d}$DESY, Hamburg, Germany,
$^{e}$SLAC, Menlo Park, CA, USA,
$^{f}$University College London, London, UK,
$^{g}$Centro de Investigacion en Computacion - IPN, Mexico City, Mexico,
$^{h}$ECFM, Universidad Autonoma de Sinaloa, Culiac\'an, Mexico
and
$^{i}$Universidade Estadual Paulista, S\~ao Paulo, Brazil.
}} \noaffiliation
\vskip 0.25cm

%% file: acknowledgement.tex
%
We thank the staffs at Fermilab and collaborating institutions,
and acknowledge support from the
DOE and NSF (USA);
CEA and CNRS/IN2P3 (France);
MON, NRC KI and RFBR (Russia);
CNPq, FAPERJ, FAPESP and FUNDUNESP (Brazil);
DAE and DST (India);
Colciencias (Colombia);
CONACyT (Mexico);
NRF (Korea);
FOM (The Netherlands);
STFC and the Royal Society (United Kingdom);
MSMT and GACR (Czech Republic);
BMBF and DFG (Germany);
SFI (Ireland);
The Swedish Research Council (Sweden);
and
CAS and CNSF (China).
Special thanks as well to John Campbell for his support with \textsc{mcfm} and Patrick Fox with \textsc{madgraph}.

%% file: plb.bbl
\begin{thebibliography}{99}
\bibitem{coord} We use a standard right-handed coordinate system. The nominal collision point is the center of the detector with
                  coordinate $(0,0,0)$. The direction of the proton beam is the positive $+z$ axis. The $+x$ axis is horizontal, pointing away from the center of the Tevatron ring. The $+y$ axis points vertically upwards. The polar angle, $\theta$, is defined ﬁned such that $\theta = 0 $ is the $+z$ direction. The rapidity is defined as $y = -\ln \left[ (E+p_Z)/(E-p_Z)\right]$, where $E$ is the energy and $p_Z$ is the momentum component along the proton beam direction. Pseudorapidity is defined as $\eta = - \ln(\tan \frac{\theta}{2})$. $\varphi$ is defined as the azimuthal angle in the plane transverse to the proton beam direction.
\bibitem{mtw} Since the longitudinal component of the momentum of the neutrinos is not measured, the measured properties of the $W$ boson candidates are limited to their transverse energy and transverse mass, defined as $M_T=\sqrt{ (\ptmiss + p_T^\ell)^2 - (\pslash_x\! + p_x^\ell)^2  - (\pslash_y\! + p_y^\ell)^2}$ where \ptmiss is the magnitude of the missing transverse energy vector, $p_T^\ell$ is the transverse momentum of the lepton and $p_x^\ell$ and $p_y^\ell$ ($\pslash_x\!$ and $\pslash_y\!$) are the magnitude of the $x$ and $y$ components of the lepton's momentum (missing transverse energy) respectively.
\end{thebibliography}
